%% file: main.tex
\documentclass[manuscript, nonacm]{acmart}
\usepackage{array}

\usepackage{xcolor}

\AtBeginDocument{%
  }

\setcopyright{none}
\renewcommand\footnotetextcopyrightpermission[1]{}

\makeatletter
\def\@mktitle@i{\hsize=\textwidth
  \setbox\mktitle@bx=\vbox{\centering\@titlefont\@title\par\bigskip}}
\def\@mkauthors@i{%
  \global\setbox\mktitle@bx=\vbox{\unvbox\mktitle@bx\par
    \vspace{6pt}\centering\normalfont
    {\Large Md Arif Shahriar, Mehedi Hasan Raju, Oleg V. Komogortsev\par}
    \vspace{3pt}
    {\large Texas State University, San Marcos, Texas, USA\par}
    \vspace{3pt}
    {\normalsize\ttfamily \{arif.shahriar, m.raju, ok\}@txstate.edu\par}
    \bigskip}}
\makeatother
\authorsaddresses{}
\acmISBN{978-1-4503-XXXX-X/2018/06}

\begin{document}

\title[Measurement-Guided Conversion for VR Authentication]{Measurement-Guided Conversion of Laboratory Eye-Tracking Data for VR Authentication}


\author{Md Arif Shahriar}
\email{arif.shahriar@txstate.edu}
\orcid{0009-0005-5133-6123}

\author{Mehedi Hasan Raju}
\email{m.raju@txstate.edu}
\orcid{0000-0002-1144-6118}

\author{Oleg Komogortsev}
\email{ok@txstate.edu}
\orcid{0000-0001-7890-8842}

\affiliation{
  \institution{Texas State University}
  \city{San Marcos}
  \state{Texas}
  \country{USA}
}

\renewcommand{\shortauthors}{Shahriar et al.}


\begin{abstract}
Eye-movement authentication in virtual reality (VR) needs longitudinal training data, yet most such datasets come from lab trackers. We convert lab recordings to a VR headset's signal quality while keeping their eye movement, using only unpaired VR recordings of other people. Measurements show that the headset's added noise dominates above 7.8~Hz and that its quality is consistent within a person. A parametric device model re-creates the headset's sampling, a person-consistent calibration error and spectrally shaped noise, and a frequency-locked diffusion renderer re-textures that high-frequency noise. Converted GazeBase recordings match GazeBaseVR in noise level and accuracy, keep their source's saccade dynamics, and are harder to tell from real VR data than two prior conversions. An authenticator trained only on converted data verifies unseen VR users at a 5.9\,\% equal error rate with 45\,s of gaze, close to real VR training (5.8\,\%); with 5\,s, a gap remains.
\end{abstract}




\maketitle

\input{sections/01_introduction}

\input{sections/02_related_work}
\input{sections/03_data_device_gap}
\input{sections/04_method}
\input{sections/05_experimental_setup}

\input{sections/06_results}
\input{sections/07_discussion}

\input{sections/08_conclusion}
\input{sections/09_statements}


\bibliographystyle{ACM-Reference-Format}
\bibliography{references}

\end{document}

%% file: sections/01_introduction.tex
\section{Introduction}
\label{sec:intro}

Eye trackers are now increasingly integrated virtual reality (VR) headsets, where they support applications such as foveated rendering, gaze interaction and user authentication \citep{clay2019vr,adhanom2023review}. 
Eye-movement biometrics, in particular, could authenticate VR users continuously and without additional hardware \citep{lohr2022ekyt,raju2024vr,lohr2024vrbaseline}. Deep authenticators such as Eye Know You Too (EKYT) \citep{lohr2022ekyt} learn identity from recordings collected across people and sessions. Training them therefore requires longitudinal datasets. Such longitudinal datasets are slow and costly to collect, and many of the largest were recorded in the lab with high-end trackers. GazeBase, for instance, follows 322 people over 37 months with an EyeLink 1000 at 1000\,Hz \citep{griffith2021gazebase}. Repeating such a collection for every new headset is impractical. Transforming existing lab recordings so that they resemble recordings of a target headset could make them usable for that headset without new data collection.

Every eye tracker produces a signal of its own quality, described by standard measures such as spatial accuracy, spatial precision, sampling rate and data loss \citep{holmqvist2012quality}. VR headset trackers record at a much lower quality than lab trackers \citep{lohr2019smi,schuetz2022vive}. Their signals also carry device-specific patterns that these measures do not capture, such as the colour of the noise \citep{niehorster2020noise}. These differences matter for every model learned from eye-tracking data. When the training and test data come from different devices, a model faces a domain shift: the features it learned are distributed differently at test time \citep{lohr2025styletransfer}. A model trained on clean lab signals can learn details that the headset blurs or never delivers. A reliable conversion must therefore reproduce every device effect, including those that no single quality measure describes, while leaving the eye movement unchanged. The resulting signals can then be evaluated in downstream applications, including biometric authentication, to assess their practical utility. It must also work without recordings of the same people on both devices, because such paired data are rarely available.

Two approaches have transformed GazeBase toward its VR counterpart, GazeBaseVR \citep{lohr2023gazebasevr}. Percentile matching \citep{griffith2025percentile} degrades the spatial accuracy, spatial precision and timing of each GazeBase recording. Each degradation places the recording at the same percentile of the GazeBaseVR quality distribution. Its added noise, however, is white, and each recording is degraded on its own. Device-specific style transfer \citep{lohr2025styletransfer} learns the transformation from unpaired recordings with MUNIT \citep{huang2018munit}. Its output looks like VR data, but the generated gaze lands far from the targets. Its output looks like VR data, but the generated mean accuracy is 20.76\textdegree that lands far from the targets \citep{lohr2025styletransfer}. Both approaches are fitted to recordings of the target device; style transfer, for instance, is trained on 100 recordings per device. Neither approach both reproduces the device's signal, including the spectrum of its noise, and keeps the eye movement of the source. It remains unclear whether a conversion can do both when it learns the device only from unpaired recordings.

\begin{figure}[htbp]
  \centering
  \includegraphics[width=0.8\columnwidth]{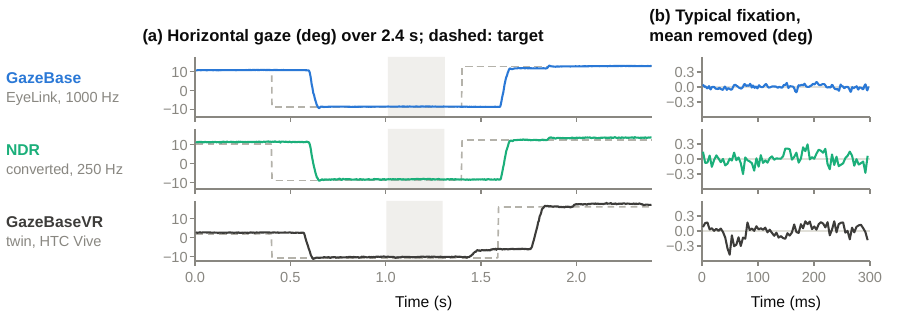}
  \caption{Horizontal gaze of one GazeBase recording before and after conversion (NDR), and a GazeBaseVR recording of its twin, a different person. (a)~2.4\,s around one fixation; dashed: target. NDR keeps the saccades of the source. (b)~The grey 300\,ms of (a), mean removed and magnified: NDR replaces the small GazeBase noise with noise of about the size of the GazeBaseVR. Each recording shows its typical fixation (the saccade-free fixation with the median spread).} 
  \label{fig:teaser}
\end{figure}

In this work, as in \citet{lohr2025styletransfer}, we assume unpaired recordings from two devices: a source with higher signal quality (the lab tracker) and a target with lower signal quality (the VR headset). In this context, we ask the following research question:
\begin{quote}
  \itshape
  Can source recordings be converted to the signal quality of the target device while keeping their eye movement?
\end{quote}

To answer it, we render the device and keep the eye movement. We model a VR recording as the eye movement plus what the device adds to it: camera sampling, a calibration error, noise and data loss. We convert a lab recording by keeping its eye movement and re-creating only these device terms. The conversion is driven by measurements: we first measure where the two devices differ (Section~\ref{sec:data}, Table~\ref{tab:gap}) and find two facts that shape the design. First, from $f^* = 7.8$\,Hz upward, the noise that the VR device adds is stronger than the lab tracker's fixation spectrum. Above this frequency, device noise dominates the VR signal. In addition, we examine how signal quality varies between participants and across sessions. These findings inform the conversion design and are evaluated in detail in Section~\ref{sec:data}, Figure~\ref{fig:spectra}.

Each conversion step then closes one measured gap that we identified. For example, Part (a) Parametric device model: First, It samples the signal as the VR camera would, with real VR timing jitter. Second, matches each lab participant by optimal transport to one VR ``twin'' of the same relative quality. Third, Replaces the recording's own calibration error with the twin's, adds device noise shaped by the twin's spectrum. Again, for Part (b) Frequency-locked diffusion renderer: a diffusion model trained only on VR recordings re-textures the noise there and leaves the signal below $f^*$ essentially unchanged. Together with Parametric device model and Frequency-locked diffusion renderer we call it neural device renderer (NDR). To evaluate the usefulness of the converted signals in a downstream task, we train an authenticator on the converted recordings and test it on unseen VR users. Finally, Figure~\ref{fig:teaser} illustrates the effect of the conversion by showing one recording before and after conversion.

Our contributions are:
\begin{enumerate}
\item A measurement of how a VR headset tracker differs from a lab tracker, not only in standard quality measures but also in the spectrum of its noise and in how consistent its quality is within a person. These measurements guide the design of the conversion.
 
\item A conversion guided by these measurements that keeps the source eye movement and re-creates the device: camera sampling, a person-consistent VR profile, calibration error and spectrally shaped noise, refined by a frequency-locked diffusion renderer. The device is learned only from unpaired VR recordings of other people.

\item An empirical evaluation on 101 unseen real VR users. It covers signal quality, realism, preservation of the eye movement and the two prior conversion methods, with authentication as a downstream test.

\end{enumerate}

%% file: sections/02_related_work.tex
\section{Related Work}
\label{sec:related}

\subsection{Eye-tracking data quality across devices and people}
Signal quality has been a concern since the early days of eye-movement research \citep{mcconkie1981quality}. \citet{holmqvist2012quality} defined accuracy and precision as the core measures of data quality and showed how to measure them, and current reporting guidelines still rest on these measures \citep{dunn2024reporting}. Data quality depends on the device: trackers recording the same eyes at the same time differ in accuracy and precision \citep{ehinger2019battery}, and commercial headsets differ in latency \citep{stein2021latency}. It also depends on the person and the procedure: calibration method, eye physiology, ethnicity and experimental design all change it \citep{nystrom2013calibration,blignaut2014ethnicity}, and artificial eyes have been studied as a way to measure a tracker's precision without a human observer \citep{wang2017artificial}. The temporal structure of the noise matters as well. \citet{niehorster2020noise} introduced measures of signal type and magnitude based on the RMS sample-to-sample deviation and the standard deviation, found that five eye trackers produce fixation signals with different characteristics, and proposed a method to generate tracker noise of any signal type and magnitude. \citet{niehorster2021drift} asked whether the spectral colour of such signals reflects fixational eye movements or the trackers' filters, and \citet{raju2021determining} determined which frequencies of eye-tracking signals carry signal and which carry noise. Trackers built into VR headsets add their own characteristics \citep{clay2019vr,adhanom2023review}: data quality has been measured for the SMI tracker in the HTC Vive \citep{lohr2019smi} and for the HTC Vive Pro Eye \citep{schuetz2022vive,sipatchin2021vive}, saccade measurement in head-mounted displays has been assessed \citep{imaoka2020saccades}, and noise has been estimated for head-mounted binocular trackers \citep{velisar2024pupilcore}. These studies measure and describe data quality; we use such measurements to convert data from one device to another. Each step of our parametric device model reproduces one measured difference between the two devices (Section~\ref{sec:gap}). We also add two measurements that the conversion needs: the frequency above which the noise the VR device adds is stronger than the lab tracker's fixation spectrum (Section~\ref{sec:fstar}), and the share of each quality measure that differs between people rather than between sessions (Section~\ref{sec:icc}).

\subsection{Synthetic eye-movement data and cross-device conversion}
Synthetic eye movements have been produced with procedural models \citep{duchowski2016synthesis}, generative adversarial networks \citep{prasse2023speyegan} and, recently, diffusion models that personalise gaze \citep{jiao2024diffeyesyn,hasan2026eyemakeyou}, and recordings degraded with additive white noise have been used to train event detectors \citep{zemblys2018ml}. Converting existing recordings to the quality of a specific device is closer to our goal. \citet{griffith2025percentile} degrade EyeLink 1000 recordings toward the SMI tracker in VR by percentile matching, on the same two dataset. \citet{lohr2025styletransfer} learned the conversion between the same datasets with MUNIT \citep{huang2018munit}; the generated signals took on the target device's precision but lost their spatial accuracy. As imaging models camera noise explicitly instead of translating whole images \citep{foi2008poissonian,brooks2019unprocessing,abdelhamed2019noiseflow},  We instead model the eye tracker, as imaging models camera noise explicitly instead of translating whole images \citep{foi2008poissonian,brooks2019unprocessing,abdelhamed2019noiseflow}. Each of our steps adds or replaces one device term on top of the source recording, and the diffusion renderer changes only a residual. The gaze therefore keeps the source's fixation positions and saccades, and its spatial accuracy changes only through the calibration error that we replace explicitly.


\subsection{Diffusion models for editing signals}
Diffusion models learn to reverse a gradual noising process \citep{sohldickstein2015,ho2020ddpm,song2021sde}. To generate a new sample, they start from pure noise and remove it step by step. The same models can also edit an existing signal instead of generating one from scratch. SDEdit \citep{meng2022sdedit} adds only part of the noise to the input and runs only the remaining denoising steps, so the coarse structure of the input survives while its fine detail is regenerated to resemble the training data. ILVR \citep{choi2021ilvr} gives tighter control: after every denoising step, it replaces the low frequencies of the sample with those of a noised reference, so the output keeps the coarse structure of that reference. Other methods shape the added noise or its schedule by frequency \citep{rissanen2023heat,davidson2026colored,esteves2026spectral}, so that the model acts more strongly on some frequencies than on others. Our renderer combines these three ideas. Like SDEdit, it starts from a partly noised version of the recording; like ILVR, it resets the low band of the sample after every step; and its noise is shaped by the spectrum of the device. It follows the EDM formulation \citep{karras2022edm} with transformer blocks in its backbone \citep{peebles2023dit}. Unlike diffusion models that synthesise gaze, it edits existing recordings rather than generating new people

\subsection{Eye-movement biometrics as a downstream task}
Eye movements have been studied as a biometric for two decades \citep{kasprowski2004,rigas2017bioeye}, and deep networks now learn identity directly from the gaze signal \citep{jager2020deepeyedentification,lohr2022ekyt}. We use EKYT \citep{lohr2022ekyt}, developed on GazeBase, as our downstream test. On GazeBaseVR, \citet{raju2024vr} reported the random-saccade baseline and \citet{lohr2024vrbaseline,lohr2026factors} studied VR authentication at scale and how signal quality influences it. Our goal is not to develop a new authentication method, but to evaluate whether converted laboratory recordings provide useful training data.

%% file: sections/03_data_device_gap.tex
\section{Data and the Device Gap}
\label{sec:data}

\subsection{Datasets and split}
\label{sec:datasets}

\input{tables/tab_datasets}
We convert the random-saccade task (RAN) of GazeBase \citep{griffith2021gazebase} to the signal quality of GazeBaseVR \citep{lohr2023gazebasevr}.
GazeBase was recorded with an EyeLink 1000 at 1000\,Hz in nine rounds over 37 months, and we use all 1,762 of its RAN recordings, from 322 people. GazeBaseVR was recorded with an SMI eye tracker inside an HTC Vive at a nominal 250\,Hz in three rounds over 26 months, while participants performed five tasks: vergence (VRG), smooth pursuit (PUR), video viewing (VID), reading (TEX) and random saccades (RAN). We use the left eye of all 5,020 recordings, from 407 people. In both datasets the RAN target jumps to random positions about every second, within $\pm15^\circ$ horizontally where $\pm9^\circ$ and $\pm10^\circ$ vertically for Gazebase and GazebaseVR respectively. Gazebase maintains each target position for 1 s, but the VR targets stay longer about 1--1.5\,s \citealp{griffith2025percentile}, so a recording holds 80 instead of 100 target positions. We split the GazeBaseVR participants by person, into set A containing 306 people (75\%); 3,780 recordings, 756 of them RAN and set B containing 101 people (25\%); 1,240 recordings, 248 of them RAN. All conversion components are developed using GazeBase and Set A RAN data, except the frequency-locked diffusion renderer, which uses all five tasks. Set B RAN is reserved exclusively for final evaluation.

\subsection{What separates the two devices}
\label{sec:gap}
Table ~\ref{tab:gap} compares resampled GazeBase (Section~\ref{sec:step1}), with the RAN recordings of GazeBaseVR set A. The two cohorts have the same median recording-level saccade latency (228\,ms). On the VR device the sample-to-sample noise is five times stronger (RMS-S2S 0.156\textdegree{} vs.\ 0.029\textdegree) and much whiter (signal type 0.780 vs.\ 0.208; spectral slope $\alpha$ 0.635 vs.\ 2.273), and gaze spreads twice as much within a fixation (MAD precision 0.175\textdegree{} vs.\ 0.086\textdegree). For all of these properties the distributions barely overlap (KS distance 0.56--0.99). In contrast, VR recordings have fewer missing samples than EyeLink recordings (median 0.4\% vs.\ 1.5\%). Because the conversion preserves source gaps, converted recordings retain the higher missing-sample rate of GazeBase. The final column of Table~\ref{tab:gap} identifies which conversion step addresses each remaining device-specific gap.

\input{tables/tab_device_gap}

\subsection{Where the added device noise takes over}
\label{sec:fstar}

\begin{figure}[htbp]
  \centering
  \includegraphics[width=0.7\linewidth]{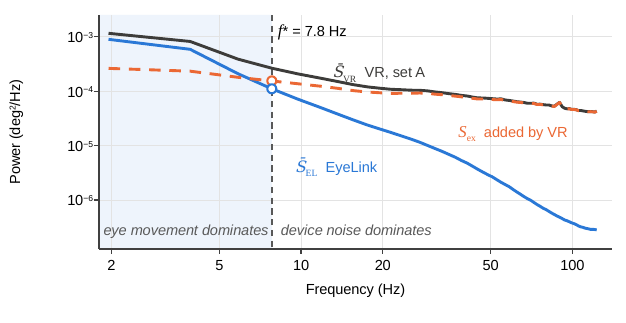}
\caption{Fixation-noise spectra of the two devices. Solid: median spectra of the GazeBase recordings (after resampling to 250\,Hz) and of the GazeBaseVR set A RAN recordings. Dashed: their difference, the noise the GazeBaseVR device adds. The added noise overtakes the GazeBase spectrum at $f^*=7.8$\,Hz (circles); in the shaded band below it, the GazeBase spectrum, mostly the eye's own movement, dominates.}
  \Description{Log-log plot of power against frequency from 2 to 123 hertz. The EyeLink spectrum falls
    steeply from about 1e-3 to 3e-7 square degrees per hertz; the VR spectrum is flatter and ends near
    4e-5. The dashed added-noise curve starts below the EyeLink curve, crosses it just before a dashed
    vertical line at 7.8 hertz and then almost coincides with the VR curve. The band left of the line is
    shaded and labelled eye movement dominates; the region right of it is labelled device noise dominates.}
  \label{fig:spectra}
\end{figure}

The noise during fixation is not white, so we compare the devices frequency by frequency. For every recording we take fixation segments of 0.5\,s, found by a velocity threshold \citep{salvucci2000ivt} on Savitzky--Golay velocity \citep{savitzky1964}. We then estimate their power spectrum at 65 frequencies between 0 and 125\,Hz (Section~\ref{sec:signature}). The median spectrum over all GazeBase recordings, $\bar S_{\mathrm{EL}}(f)$, contains the eye's fixational movement, including drift, and EyeLink's own, small noise \cite{raju2021determining}. Similarly, the median over the 756 GazeBaseVR set A RAN recordings, $\bar S_{\mathrm{VR}}(f)$, contains eye movement plus the VR device's noise. Assuming comparable fixation behavior across the two participant pools, their difference, $S_{\mathrm{ex}}(f) = \bar S_{\mathrm{VR}}(f) - \bar S_{\mathrm{EL}}(f)$, is the noise the VR device adds beyond the EyeLink's (Figure~\ref{fig:spectra}). We define $f^*$ as the lowest frequency, from 2\,Hz upward, at which this added noise is at least as strong as the EyeLink spectrum, $S_{\mathrm{ex}}(f) \ge \bar S_{\mathrm{EL}}(f)$. With spectra spaced 1.95 Hz apart, that gives $f^* = 7.8$\,Hz. Below $f^*$ the added noise is weaker than the EyeLink spectrum but not absent; above it, it dominates. $f^*$ is therefore an empirical crossover of these two datasets under the assumption that both groups fixate alike, and we use it as a design choice for the renderer, not as a physiological boundary. It also does not separate identity information from noise, as eye-tracking signals retain identity information in their high-frequency components \citep{raju2024signal}. Parametric device model (Step~4) adds device noise at all frequencies; the learned renderer leaves the band below $f^*$ essentially unchanged and re-textures the band above it, where the estimated excess noise is stronger (Section~\ref{sec:renderer}).

\subsection{Device quality is consistent within a person}
\label{sec:icc}
Device quality also differs between people. For each property we compute the intraclass correlation Coefficient ICC(1) \citep{shrout1979icc} over each participant's RAN recordings. Following \citet{friedman2017persistence}, we interpret ICC as a measure of temporal persistence: the proportion of variance attributable to differences between people rather than between sessions. On the GazeBaseVR device, 82\,\% of the variance of the noise level, 73\,\% of that of precision and 75\,\% of that of the signal type lie between people (Table~\ref{tab:gap}). A set A participant has 2.5 RAN recordings on average, mostly the two contiguous sessions of one visit \citep{lohr2023gazebasevr}, so these values describe consistency within a visit more than over months; the 5.5 GazeBase recordings of a participant span up to 37 months. A participant who is noisy in one GazeBaseVR session is therefore noisy in the next. A conversion that drew a random GazeBaseVR profile for every recording would give each converted person a different noise level in every session, a structure that real GazeBaseVR data do not show. Parametric device model (Step~2) therefore assigns each GazeBase participant one GazeBaseVR device profile, while keeping each session's own variation (Section~\ref{sec:step2}).

%% file: tables/tab_datasets.tex
\begin{table}[t]
  \caption{Datasets used in our study. GazeBaseVR participants are split by person into Set A for training and Set B for testing. All five tasks are used for Set A; only RAN is used for Set B.}
  \label{tab:datasets}
  \small
  \begin{tabular}{lll}
    \toprule
                  & GazeBase (source)            & GazeBaseVR (target) \\
    \midrule
    Device        & EyeLink 1000                 & SMI eye tracker in an HTC Vive \\
    Sampling      & 1000\,Hz                     & 250\,Hz \\
    Tasks used    & RAN                          & VRG, PUR, VID, TEX, RAN \\
    Recordings / people & 1,762 / 322            & 5,020 / 407 \\
    Set A (train) & --                           & 3,780 recordings (all five tasks) / 306 people \\
    Set B (test) & --                            & 248 RAN recordings / 101 people \\
    Eye used      & left                         & left \\
    \bottomrule
  \end{tabular}
\end{table}

%% file: tables/tab_device_gap.tex

\begin{table}[htbp]
\caption{What separates the devices after resampling to 250\,Hz. Medians are computed over recordings (GazeBase: 1,762 RAN recordings; GazeBaseVR: 756 RAN recordings from Set A). KS measures the distance between distributions (0 = identical). Person share is ICC(1), the proportion of variance attributable to differences between people. Step 4 denotes the parametric device model, and the renderer denotes the frequency-locked diffusion renderer. See Figure~\ref{fig:pipeline} for details.}
  \label{tab:gap}
  \small
  \begin{tabular}{lrrrrrl}
    \toprule
                         & \multicolumn{2}{c}{Median} &      & \multicolumn{2}{c}{Person share} & \\
    \cmidrule(lr){2-3} \cmidrule(lr){5-6}
    Property             & EyeLink & VR    & KS   & EyeLink & VR   & Closed by \\
    \midrule
    Accuracy        & 0.860   & 1.407 & 0.56 & 0.37    & 0.28 & Steps 2--3 \\
    MAD precision   & 0.086   & 0.175 & 0.70 & 0.40    & 0.73 & Step 4, renderer \\
    RMS-S2S         & 0.029   & 0.156 & 0.99 & 0.68    & 0.82 & Step 4, renderer \\
    STD             & 0.134   & 0.202 & 0.61 & 0.45    & 0.72 & Step 4, renderer \\
    Signal type          & 0.208   & 0.780 & 0.98 & 0.75    & 0.75 & Step 4, renderer \\
    Slope $\alpha$       & 2.273   & 0.635 & 0.98 & 0.67    & 0.67 & Step 4, renderer \\
    Data loss (\%)       & 1.5     & 0.4   & 0.37 & 0.44    & 0.44 & source gaps kept \\
    \midrule
    Saccade latency (ms) & 228     & 228   & 0.06 & 0.64    & 0.87 & nothing to close \\
    \bottomrule
  \end{tabular}
\end{table}

%% file: sections/04_method.tex

\section{Method}
\label{sec:method}

\subsection{Overview: render the device, keep the eye movement}
\label{sec:overview}
We model a GazeBaseVR recording as the eye movement plus what the device adds to it:
\begin{equation}
  g_{\mathrm{VR}}(t) = \mathcal{S}[e](t) + c_{\mathrm{VR}}\big(e(t)\big) + n_{\mathrm{VR}}(t),
  \label{eq:model}
\end{equation}

\begin{figure}[htbp]
  \centering
  \includegraphics[width=0.8\linewidth]{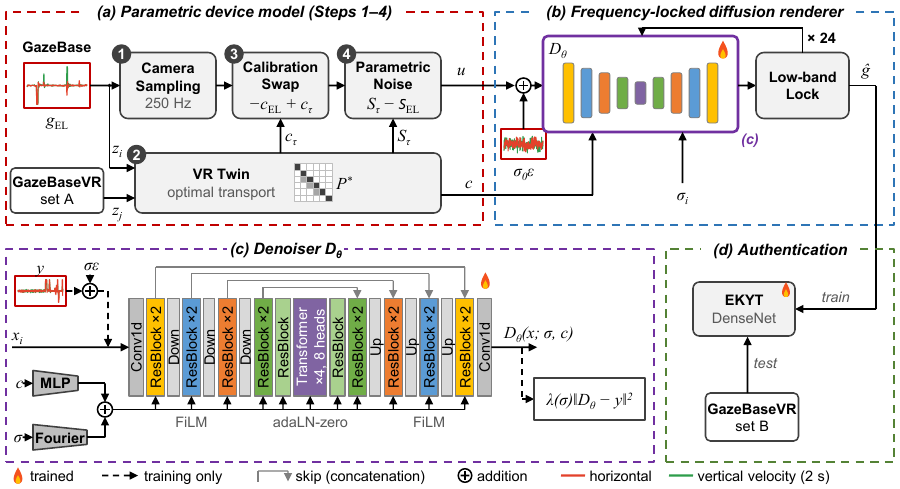}
  \caption{Overview of the proposed modela and its evaluation. It consist of four parts, First, \textbf{(a)}~Parametric device model
    (Steps~1--4), which turns a GazeBase recording $g_{\mathrm{EL}}$ into the proposal $u$. Second, \textbf{(b)}~Frequency-locked diffusion renderer, which turns $u$ into $\hat g$ and keeps the band below $f^* = 7.8$\,Hz essentially unchanged. Third, \textbf{(c)}~The denoiser $D_\theta$ trained on set A on all five tasks. Finally, \textbf{(d)}~Authentication with EKYT on the unseen set B RAN task. Section~\ref{sec:overview} describes each part of the model.}
  \label{fig:pipeline}
\end{figure}

Equation~\ref{eq:model} describes the eye movement $e(t)$, the camera's sampling $\mathcal{S}$, the calibration error $c_{\mathrm{VR}}$, and the device noise $n_{\mathrm{VR}}$. It does not explicitly model device filtering: a real tracker may also smooth or delay eye movements through its exposure, gaze estimation, and internal filtering. We model the camera's short exposure but do not include a separate filter $H$, because separating device filtering from differences between the participant groups would require recordings of the same people on both devices. A lab recording from GazeBase has the same general form, but with different sampling, smaller calibration error, and much lower noise (Section~\ref{sec:gap}). We therefore retain the eye movements in GazeBase and replace the device effects that our pipeline models. Figure~\ref{fig:pipeline} shows the conversion pipeline where \textbf{(a)}~The parametric device model takes a GazeBase recording $g_{\mathrm{EL}}$ (1000\,Hz) and returns the \emph{parametric proposal} $u$. \textbf{(b)}~The frequency-locked renderer perturbs $u$ with shaped noise and refines it over 24 steps using the pretrained denoiser $D_\theta$, resetting the band below $f^*$ to that of $u$ after each step. It returns $\hat g = u + r_\theta$, where the residual $r_\theta$ modifies essentially only the noise texture above $f^*$. \textbf{(c)}~The denoiser $D_\theta$ is a diffusion model trained on Set~A before conversion. it takes the noise level $\sigma$ and conditioned on the twin's texture features $c$, to a clean estimate. \textbf{(d)}~For evaluation, EKYT is trained on the converted recordings $\hat g$ and tested on recordings from the unseen GazeBaseVR participants in set~B RAN task.

\subsection{Part (a): Parametric device model}
\subsubsection{\textbf{Step 1: camera sampling}}
\label{sec:step1}
A camera integrates light over a short exposure and, in the VR headset, takes its samples at slightly irregular instants. We reproduce both. The SMI camera's exposure time is not reported in the available studies \citep{lohr2023gazebasevr,lohr2019smi}. We therefore approximate it with a 2,ms exposure window, half the 4,ms frame period. Camera instants are then drawn as $t_0 = 0$ and $t_{k+1} = t_k + \Delta_k$, with each interval $\Delta_k$ drawn independently from the 4.27 million inter-sample intervals of GazeBaseVR recordings (mean 4.00\,ms, SD 0.19\,ms). Real sampling intervals may be temporally correlated, but we draw them independently from the observed GazeBaseVR intervals. Then, the averaged signal is read at each instant $t_k$ by linear interpolation between the two neighbouring 1\,ms samples, and the irregular samples are resampled onto an exact 250\,Hz grid with PCHIP interpolation \citep{fritsch1980pchip}, as are the GazeBaseVR recordings. After this resampling, the jitter no longer changes the sampling times; it leaves only a small interpolation error, largest where the gaze moves fast. It changes the signal by about 0.004\textdegree{} RMS during fixations, against a VR noise level of 0.156\textdegree, and by about 0.02\textdegree{} over whole recordings, mostly during saccades. The samples those were Missing stay missing.

\textbf{The device signature}
\label{sec:signature}
Every recording of both devices is described by the same measurements, its \emph{signature}. Each GazeBase recording (all RAN) receives the full signature. On GazeBaseVR, the texture part needs no target and is measured on all five tasks; the geometric part needs the target and is measured on RAN only.
\begin{itemize}
  \item \emph{Texture.} Fixation pieces of 0.5\,s are taken where the Savitzky--Golay speed (window 7, order 2) stays below 30\textdegree/s \citep{salvucci2000ivt, savitzky1964}. For each piece and channel we remove the linear trend, apply a Hann window and compute the power spectrum with a 128-point FFT, which gives 65 frequencies from 0 to 125\,Hz; the spectra are averaged over pieces and channels. From the pieces we also take the median RMS sample-to-sample deviation (RMS-S2S) and standard deviation (STD), their ratio, the signal type, which is $\sqrt 2$ for white noise \citep{niehorster2020noise}, the slope $\alpha$ of $S(f) \propto f^{-\alpha}$ between 4 and 100\,Hz, and the log power in six octave bands from 2 to 125\,Hz.
  \item \emph{Geometry.} Saccade latency is the shift that best aligns gaze with the target. With gaze shifted by this latency, fixations are the 500\,ms that follow the first 400\,ms after each target jump. Outlier removed using Tukey’s fences (1.5 times the interquartile range beyond either quartile) or excluding samples with gaze error exceeding 2\textdegree. Accuracy is calculated using the mean distance between gaze and target (for GazeBaseVR, the target as seen from the recorded eye, not from the cyclopean eye) and precision the MAD about the geometric median \citep{lohr2019smi}. The calibration error of each channel is described by the quadratic field $c(x, y) = A x^2 + B y^2 + C x + D y + E$ \citep{lohr2019smi}, over the target position $(x, y)$. We fit the field by least squares to the offsets of fixation centroids from their targets, using a ridge term of $10^{-3}$ on all coefficients except $E$, and at least 8 fixations.
  \item \emph{Data loss.} The fraction of missing samples and the mean length of a gap.
\end{itemize}
Altogether 89 profile number is created after Texture and Geometry transformation where Four noise numbers, six band powers, 65 spectrum values, ten field coefficients, accuracy, precision and two data-loss are present.

\subsubsection{\textbf{Step 2: a person-consistent VR profile (the twin)}}
\label{sec:step2}
Because GazeBaseVR quality is largely consistent of the person (Section~\ref{sec:icc}), each GazeBase participant needs one GazeBaseVR participant, its twin. We describe every participant by the means, over their RAN recordings, of five signature features: log accuracy, log precision, log RMS-S2S, signal type and slope $\alpha$. Participants are compared by their relative standing: the features are standardised within each dataset, giving vectors $z_i$ for the 322 GazeBase participants and $z_j$ for 305 of the 306 set A participants; the remaining one has no 0.5\,s fixation piece below 30\textdegree/s in either RAN recording, so its noise features are undefined. We then match the two populations by solving, as a linear program, the exact optimal-transport problem \citep{peyre2019ot}
\begin{equation}
  P^\star = \operatorname*{arg\,min}_{P \in \mathcal{U}(n,m)} \; \sum_{i,j} P_{ij}\, C_{ij},
  \label{eq:ot}
\end{equation}
where $C_{ij} = \tfrac{1}{5}\lVert z_i - z_j \rVert^2$ is the cost of pairing GazeBase participant $i$ with GazeBaseVR participant $j$, and $\mathcal{U}(n,m)$ contains all couplings $P \ge 0$ with $P\mathbf{1} = \tfrac{1}{n}\mathbf{1}$ and $P^{\top}\mathbf{1} = \tfrac{1}{m}\mathbf{1}$: each of the $n = 322$ GazeBase participants sends the same mass, and each of the $m = 305$ GazeBaseVR participants receives the same mass. We take as twin $\tau(i) = \arg\max_j P^\star_{ij}$, which receives a median 94\,\% of participant $i$'s mass. Unlike nearest-neighbour matching, the uniform marginals spread the GazeBase participants over the whole GazeBaseVR population: all 305 GazeBaseVR participants are used, 288 once and 17 twice. In one dimension this coupling reduces to percentile matching \citep{griffith2025percentile}. Twins are close (mean cost 0.22, against 1.99 for random pairs) and preserve the participants' order (Spearman $\rho$ 0.85--0.94 between participants' and twins' features).

Each GazeBase recording $r$ of participant $i$ then receives its own profile of 89 numbers. For each number, the twin's mean is shifted by the recording's deviation from its participant's mean, mapped by quantile from the GazeBase distribution of within-person deviations to the VR one:
\begin{equation}
  p_r = \mu^{\mathrm{VR}}_{\tau(i)} + Q_{\mathrm{VR}}\!\Big(F_{\mathrm{EL}}\big(p^{\mathrm{EL}}_r -
  \mu^{\mathrm{EL}}_i\big)\Big),
  \label{eq:profile}
\end{equation}
where $p^{\mathrm{EL}}_r$ is the recording's own signature, $\mu$ are participant means, $F_{\mathrm{EL}}$ is the distribution function of GazeBase within-person deviations and $Q_{\mathrm{VR}}$ the quantile function of GazeBaseVR within-person deviations. We call $p_r$ the recording's assigned profile. The twin's mean keeps a participant consistent across sessions; the mapped deviation keeps session-to-session variation at the size the GazeBaseVR shows.

\subsubsection{\textbf{Step 3: calibration replacement}}
The GazeBase recording already carries its own calibration error, so we replace this error with the twin's instead of adding the twin's on top of it:
\begin{equation}
  g(t) \leftarrow g(t) - c_{\mathrm{EL}}\big(g(t)\big) + c_{\tau}\big(g(t)\big),
  \label{eq:swap}
\end{equation}
where $c_{\mathrm{EL}}$ is the field fitted to the recording itself and $c_\tau$ the field in its assigned profile, both evaluated at the current gaze position, which approximates the target position during fixations. Adding the twin's field without removing the recording's own would leave two calibration errors in the signal. The fitted fields are estimates: their offsets may reflect not only tracker calibration error but also the participant's fixation behavior.

\subsubsection{\textbf{Step 4: parametric device noise}}
\label{sec:step4}
A GazeBaseVR fixation spectrum contains the eye's own movement plus the VR device's noise; a GazeBase recording also contains its eye's movement plus the device's much smaller noise. We therefore add only the difference, estimated from the assigned profile, and synthesise noise whose spectrum matches it in expectation
\citep{niehorster2020noise}:
\begin{equation}
  S_r(f) = \max\!\big(S_{\tau,r}(f) - \bar S_{\mathrm{EL}}(f),\; 0.05\, S_{\tau,r}(f)\big),
  \qquad
  n = \mathcal{F}^{-1}\!\Big[\sqrt{S_r(f)\, f_s / 2}\;\, \mathcal{F}[w](f)\Big].
  \label{eq:noise}
\end{equation}
Here $S_{\tau,r}$ is the profile's spectrum, $\bar S_{\mathrm{EL}}$ the typical EyeLink spectrum (The median spectrum over all GazeBase recordings), $w$ white Gaussian noise, $\mathcal{F}$ the Fourier transform and $f_s = 250$\,Hz; the floor keeps a minimum of noise where the two spectra are close. The twin's spectrum $S_{\tau,r}$ contains the twin's eye movement plus the VR device's noise, and the GazeBase recording already contains its own eye movement plus the EyeLink's noise. Subtracting the typical EyeLink spectrum removes a typical eye movement and the typical EyeLink noise, so what remains is approximately the extra noise of the VR device. If the twin's eye moves more than a typical eye, part of that movement stays in $S_r$ and is added as noise; if it moves less, the floor applies. These errors are limited to low frequencies, where eye movement is strong; above $f^*$ the VR noise dominates and they are negligible. However, we use the typical spectrum rather than the recording's own because the twin's eye was never recorded on the EyeLink, and the recording's own spectrum describes a different eye. The zero-frequency term is set to zero, so the calibration of Step~3 is untouched, and $n$ has the one-sided power spectrum $S_r$. Noise is drawn independently for the horizontal and vertical channels. Missing samples, such as blinks, are first filled by linear interpolation, so the noise is added to a continuous signal, and are set back to missing afterwards. Finally, if the profile's data-loss fraction is higher than the recording's own, extra gaps are inserted: their number is Poisson-distributed, their positions are random and their lengths are exponentially distributed with the profile's mean gap length. These gaps are the last operation of the conversion, so in NDR they are inserted after the renderer. The result of Step~4, before these gaps, is the parametric proposal $u$: its noise has the right level and spectrum, but it is Gaussian and the same over time.

\subsection{Part (b): Frequency-locked diffusion renderer}
\subsubsection{\textbf{Frequency-locked diffusion renderer}}
\label{sec:renderer}
The renderer Fig~\ref{fig:pipeline}b replaces the Gaussian, time-invariant texture of the proposal $u$ with real GazeBaseVR texture while keeping its low-frequency content. It uses the denoiser $D_\theta$, which is trained beforehand on set A and kept fixed here, see Fig~\ref{fig:pipeline}b and Section~\ref{sec:denoiser} denoiser training details. A recording is processed in windows of 4,096 samples with a hop of 2,048. Following SDEdit \citep{meng2022sdedit}, sampling starts from the proposal perturbed at the device level, $x_0 = u + \sigma_0 \varepsilon$ with $\sigma_0 = k\,\sigma_{\mathrm{dev}}$, where $\sigma_{\mathrm{dev}} = (\int S_r(f)\, df)^{1/2}$ is the level of the Step~4 noise and $\varepsilon$ is shaped by $S_r$, and runs through 24 decreasing noise levels down to 0.02\textdegree{} ($\rho = 7$)\citep{karras2022edm} and then to zero, with Heun steps (a plain Euler step for the last). After every step, a lock in the spirit of ILVR \citep{choi2021ilvr} replaces the low band of the sample with that of the proposal, noised to the next noise level:
\begin{equation}
  x_{i+1} \leftarrow \tilde x_{i+1} - L\big(\tilde x_{i+1}\big) + L\big(u + \sigma_{i+1}\,\varepsilon_{i+1}\big),
  \label{eq:lock}
\end{equation}
where $\tilde x_{i+1}$ denotes the Heun update, $\varepsilon_{i+1}$ is fresh noise shaped by the device noise spectrum, and $L$ is a zero-phase low-pass filter with unit response below $f^*-3$\,Hz, a raised-cosine transition between $f^*-3$\,Hz and $f^*+3$\,Hz (response 0.5 at $f^*$), and zero response above $f^*+3$\,Hz. At the final step, $\sigma=0$, so the renderer's change to a window is $(I-L)(\tilde x - u)$: it is exactly zero below $f^*-3 = 4.8$\,Hz, attenuated in the transition band, and free above $f^*+3 = 10.8$\,Hz. For each window, only the residual $x_{24}-u$ is kept. Residuals from overlapping windows are blended with Hann weights and added to the proposal, which preserves its absolute position across window boundaries. Because the blending can mix frequencies slightly, we measure the result on the assembled recordings (Section~\ref{sec:res-content}): the change stays below $10^{-11}$ of the proposal's power under 4\,Hz and below $10^{-5}$ at $f^*$. The lock keeps the low band of the proposal $u$, not of the original recording; below $f^*$, $u$ differs from the source only by the device terms that Steps~1--4 add, which are intended. Finally, the original missing-sample locations are restored, and the GazeBaseVR data-loss process is applied.

The perturbation size $k$ is chosen on set A alone, by a round trip: real set A recordings are perturbed and settled by the renderer, and an EKYT model that never saw the participant verifies each of them against the participant's untouched other session. We take the largest tested $k$ that keeps at least 95\,\% of the untouched $d'$ while changing the signal type by at most 0.10 and the log noise level by at most 0.15. This gives $k = 1$ (96.4\,\% of $d'$; $k = 2$ keeps 92.5\,\%).

\subsection{Part (c): The denoiser and its training}
\subsubsection{\textbf{The denoiser and its training}}
\label{sec:denoiser}
We train a denoiser $D_\theta(x; \sigma, c)$ (Figure~\ref{fig:pipeline}c) on GazeBaseVR set A only, on all five tasks, resulting 3,761 usable recordings. Each example is a random 16.4\,s crop (4,096 samples) with at least 90\,\% valid samples and positions divided by 10\textdegree; gaps are filled linearly in the input and excluded from the loss. As in EDM \citep{karras2022edm}, the clean crop $y$ is corrupted to $y + \sigma\varepsilon$, but the noise $\varepsilon$ is shaped: it has the spectrum of the recording's own device noise, normalised to unit variance per sample. The model is conditioned on the noise level $\sigma$ and on $c$, the recording's ten standardised texture numbers (RMS-S2S, STD, signal type, slope and six band powers), and
minimises
\begin{equation}
  \mathbb{E}_{y,\sigma,\varepsilon}\Big[\lambda(\sigma)\, \big\lVert D_\theta(y + \sigma\varepsilon;
  \sigma, c) - y \big\rVert^2_{\mathrm{valid}}\Big],
  \qquad \lambda(\sigma) = \frac{\sigma^2 + \sigma_{\mathrm{d}}^2}{(\sigma\, \sigma_{\mathrm{d}})^2},
  \label{eq:loss}
\end{equation}
where $\sigma_{\mathrm{d}}$ is the standard deviation of the data. Half of the noise levels are drawn from the EDM log-normal distribution and half log-uniformly within a factor of four of the recording's device noise level $\sigma_{\mathrm{dev}} = (\int S(f)\, df)^{1/2}$, the range used at conversion in part b. As the shaped noise has a fixed covariance, the probability-flow ODE and its Heun sampler apply unchanged. The network is a one-dimensional U-Net \citep{ronneberger2015unet} with 64, 128, 256 and 256 channels and two residual blocks per level, with four transformer layers of eight heads at its 512-sample bottleneck \citep{peebles2023dit}. The noise level (through Fourier features) and $c$ (through a small MLP, replaced by a learned null vector for 10\,\% of the examples) modulate every residual block (FiLM) and transformer layer (adaLN-zero). At noise levels near $\sigma_{\mathrm{dev}}$, the trained denoiser removes only a small fraction of the added noise, allowing the renderer to refine the signal texture while retaining device-like noise characteristics.

\subsection{Part (d): Authentication with converted training data}
\subsubsection{\textbf{Authentication with converted training data}}
\label{sec:auth}
The converted recordings $\hat g$ are meant as training data for eye-movement authentication on the target headset, so we evaluate them in that setting (Figure~\ref{fig:pipeline}d): EKYT \citep{lohr2022ekyt} is trained only on converted recordings and tested on real GazeBaseVR users (Section~\ref{sec:setup}).

%% file: sections/05_experimental_setup.tex
\section{Experimental Setup}
\label{sec:setup}

\subsection{Conditions}
We convert all 1,762 GazeBase RAN recordings under three conditions; a fourth, Real VR, uses GazeBaseVR set A recordings as the reference. All are compared on Set B RAN task.
\begin{itemize}
  \item \textbf{Plain}: camera sampling only (Step~1).
  \item \textbf{Parametric}: the parametric device model (Steps~1--4).
  \item \textbf{NDR}: the parametric device model followed by the renderer.
  \item \textbf{Real VR}: 756 RAN recordings of GazeBaseVR set A.
\end{itemize}

\subsection{Training}
The denoiser $D_\theta$ contains 12.6 million parameters and is trained exclusively on set A all tasks for 150,000 optimization steps, using batches of 32 crops. We use AdamW with a learning rate of $2\times10^{-4}$ and no weight decay, with linear warm-up over the first 2,000 steps followed by cosine decay to zero. Gradients are clipped to a norm of 1, and training uses bfloat16 mixed precision. An exponential moving average (EMA) of the model weights, with a decay rate of 0.999, is maintained and used during conversion. All experiments are implemented in Python using PyTorch 2.5.1 and pytorch-metric-learning 2.9.0. Training takes approximately 5 h on a single NVIDIA RTX 3080 GPU. Random number generators are initialized with fixed seeds to support reproducibility.

For the authenticator, we reimplement EKYT \citep{lohr2022ekyt} and train it separately for each condition, with the settings of the GazeBaseVR baseline \citep{raju2024vr}. Its input is the horizontal and vertical gaze velocity (Savitzky--Golay filter, order 2, 7 samples), clipped to $\pm 1000$\textdegree/s, z-scored with one mean and standard deviation of the training set, and cut into non-overlapping 5\,s windows; missing samples are set to zero after z-scoring, and windows with more than half of their samples missing are dropped. The network, a pre-activation DenseNet of eight dilated 1D convolutions with a 128-dimensional embedding, is trained for 100 epochs on batches of 8 participants $\times$ 8 windows, with the multi-similarity loss \citep{wang2019multisimilarity} plus 0.1 times a cross-entropy loss, and Adam with a one-cycle learning rate that rises from $10^{-4}$ to $10^{-2}$ over the first 30\,\% of training and then falls to $10^{-7}$. There is no validation set or early stopping; the final weights are used. The training participants are split at random into four disjoint folds, and one model is trained on each set of three folds.

\subsection{Signal quality, realism and content}
We compare each condition with the 248 RAN recordings of set B on the signature measures (Section~\ref{sec:signature}), using medians and the two-sample Kolmogorov--Smirnov (KS) distance. Following \citet{griffith2025percentile}, we measure realism as the accuracy of a 1-nearest-neighbour two-sample test \citep{lopezpaz2017c2st} on six standardised measures (median over five draws; 50\,\% means indistinguishable), and person share is the ICC(1) of Section~\ref{sec:icc}. To check that the eye movement is kept, we compare each converted recording with its source (number of fixations, saccade latency) and the saccade main sequence (median peak velocity per amplitude bin, 150 recordings per set).

\subsection{Authentication protocol}
Each condition's EKYT is tested on the RAN recordings of the 101 participants of set B,, whom no part of training has seen. At test time, each set B participant enrols with round~1 session~1 and is verified with round~1 session~2. The four fold models form an ensemble. Each window is embedded by all four models, and their 128-dimensional embeddings are L2-normalised and concatenated. For each side of a comparison, we average the embeddings from either the first window (5\,s) or first nine windows (45\,s), and normalise the resulting vector again. The comparison score is the cosine similarity between the two vectors. Participants with too few valid windows would be left out. We report the equal error rate (EER); the decidability index $d' = |\mu_{\mathrm{g}} - \mu_{\mathrm{i}}| / \sqrt{(\sigma_{\mathrm{g}}^2 + \sigma_{\mathrm{i}}^2)/2}$ of the genuine and impostor scores; the false rejection rate (FRR) at a false acceptance rate of $10^{-4}$; and the rank-1 identification rate. Confidence intervals (95\,\%) come from 2,000 bootstrap resamples of the test participants \citep{efron1993bootstrap}; paired differences in $d'$ between two conditions use the same resamples. An interval that contains zero shows no detectable difference, but it does not show that the two conditions are equivalent. Each condition is trained once, so the intervals cover the sampling of test participants, not training randomness.

%% file: sections/06_results.tex
\section{Results}
\label{sec:results}

\subsection{Signal quality and realism}
\label{sec:res-quality}
\input{tables/tab_quality}
Table~\ref{tab:quality} shows that the renderer brings the noise closest to that of GazeBaseVR. Its median RMS-S2S is 0.142\textdegree{} against 0.147\textdegree{} for Set B (KS 0.06, against 0.16 for parametric and 0.99 for plain data), and its signal type 0.788 against 0.770 (KS 0.12, against parametric 0.24). Accuracy reaches the Set B level in both converted conditions (KS 0.06). A nearest-neighbour classifier tells plain data from Set B data almost perfectly (99.2\,\%), parametric data less often (73.4\,\%) and NDR data less often still (70.1\,\%). The renderer's gain is a trade-off, however: it moves three properties slightly further from the VR device than the parametric data are, namely precision (KS 0.25 to 0.32), STD (0.05 to 0.11) and slope $\alpha$ (0.12 to 0.15). Precision remains too high relative to the target (MAD 0.129\textdegree{} vs.\ 0.166\textdegree): converted signals spread less within a fixation than VR signals, and the renderer cannot add slow drift because the lock keeps the slow band of the proposal. Converted data also lose more samples than Set B data (1.8\,\% vs.\ 0.3\,\%): they keep the gaps of their sources (1.5\,\%), and VR-type gaps are added only where the assigned profile has more data loss than the source recording. NDR reproduces the person share of the Set B noise level (ICC 0.80 for NDR and 0.78 for parametric data, against 0.84 for Set B).

\input{tables/tab_auth}

\subsection{Authentication of real VR users}
\label{sec:res-auth}

\begin{figure}[htbp]
  \centering
  \includegraphics[width=\linewidth]{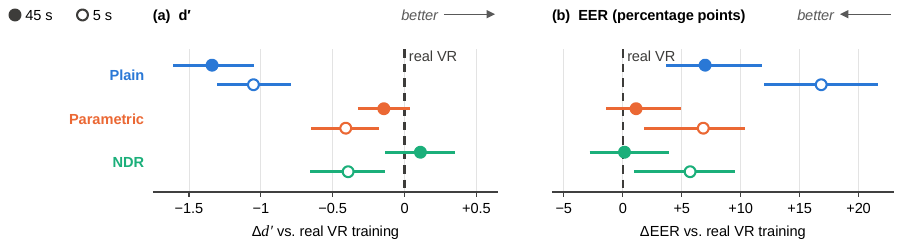}
  \caption{Authentication Set B users by EKYT trained on each converted set, relative to EKYT trained on real VR data (dashed line): difference in (a)~$d'$ and (b)~EER, with paired 95\,\% bootstrap confidence intervals, for 45\,s (filled) and 5\,s (open) of data.}
  \Description{Two forest plots side by side, one row per training set (plain, parametric, NDR), each with a 45-second and a 5-second estimate and its confidence interval, relative to a dashed
    zero line for real VR training data. Left, difference in d-prime: plain lies far below zero (about
    minus 1.3 at 45 s and minus 1.1 at 5 s); at 45 s the parametric and NDR intervals both
    include zero, with NDR at plus 0.11; at 5 s all converted sets lie below zero (about minus 0.4 to
    minus 0.5). Right, difference in equal error rate in percentage points: plain lies far above zero
    (plus 7 at 45 s, plus 17 at 5 s); at 45 s parametric and NDR include zero; at 5 s all converted sets lie above zero (plus 5 to plus 7).}
  \label{fig:results}
\end{figure}

Table~\ref{tab:auth} lists the results per condition set, and Figure~\ref{fig:results} shows each set's difference from training on real VR data. With 45\,s of data, EKYT trained only on NDR-converted GazeBase recordings reaches $d' = 3.21$ [2.97, 3.48] on unseen GazeBaseVR set B users, against 3.10 [2.90, 3.32] when trained on real VR recordings; their paired difference, $+0.11$ [$-0.14$, $+0.35$], includes zero (Figure~\ref{fig:results}a): the two are close, but the interval does not show that they are equivalent. The EERs are 5.9\,\% and 5.8\,\% (Figure~\ref{fig:results}b). The interval of the parametric set also includes zero at 45\,s ($-0.14$ [$-0.32$, $+0.04$]). Without a device model, the authenticator fails: plain resampling reaches $d' = 1.76$, $-1.34$ [$-1.61$, $-1.05$] below real VR, and its EER is 7.0 percentage points higher. The device model makes most of the difference (parametric $-$ plain: $+1.20$ [$+0.91$, $+1.45$]). Its interval against real VR also includes zero, so coming close to real VR at 45\,s does not require the renderer. The renderer still adds to it: compared directly, NDR is reliably above the parametric set ($+0.25$ [$+0.13$, $+0.38$]), and it brings the noise level and spectrum closest to those of the VR device (Section~\ref{sec:res-quality}). With 5\,s of data, every converted set stays below real VR (open markers in Figure~\ref{fig:results}; NDR: $d'$ 1.47 vs.\ 1.86), and the renderer no longer helps.

FRR at a FAR of $10^{-4}$ and rank-1 identification tell a similar story. At 45\,s, NDR rejects 64.4\,\% of genuine attempts, against 56.4\,\% for real VR and 96.0\,\% for plain data, and identifies 80.2\,\% of participants at rank 1, against 86.1\,\% and 67.3\,\%; rank-1 places the parametric set (82.2\,\%) slightly above NDR. With 101 test participants, a FAR of $10^{-4}$ is set by about the highest of 10,100 impostor scores, so this FRR is exploratory.

\subsection{The coarse eye movement is retained}
\label{sec:res-content}
\begin{figure}[htbp]
  \centering
  \includegraphics[width=\linewidth]{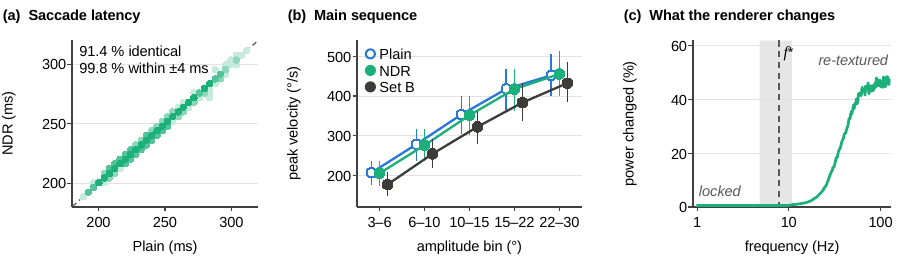}
    \caption{Preservation of coarse eye-movement characteristics. (a)~Recording-level saccade latency. (b)~Saccade main sequence: median peak velocity by amplitude bin. (c)~Relative power change introduced by the renderer; grey indicates the transition band around $f^*$.}
  \Description{Three panels. Left: a scatter plot of NDR against Plain saccade latency between 180
    and 320 milliseconds; all points lie on the diagonal, with 91.4 percent identical and 99.8 percent within
    4 milliseconds. Middle: peak velocity against saccade amplitude bin for Plain and NDR; the two curves
    coincide. Right: on a logarithmic frequency axis, the share of power changed by the renderer stays at
    zero below about 5 hertz, rises across a grey band around 7.8 hertz and levels off near 45 percent
    above 60 hertz.}
  \label{fig:content}
\end{figure}

Figure~\ref{fig:content} shows that the conversion keeps the coarse eye movement of its source. The saccade latency, measured once per recording, is unchanged in 91.4\,\% of the recordings and changes by at most one sample (4\,ms) in 99.8\,\% (Figure~\ref{fig:content}a), and the number of fixations per recording stays the same (96.7 after Plain, 96.8 after the NDR). The saccade main sequence of the converted data follows that of the plain data (Figure~\ref{fig:content}b; median peak velocity 204 vs.\ 206\textdegree/s for 3--6\textdegree{} saccades). Both, however, lie above the real VR data of set B in every amplitude bin. The conversion keeps the source's saccade dynamics and does not reproduce the slower saccades of the VR device. Figure~\ref{fig:content}c shows where the renderer acts. Left of the grey band, below 4.8\,Hz, the lock is exact and the renderer. Inside the grey band, the lock's transition from 4.8 to 10.8\,Hz around $f^*$, the change stays below 0.3\,\%. Right of the band, it rises to 16\,\% of the power at 32\,Hz and about 45\,\% above 60\,Hz, where the renderer re-textures the device noise.

\input{tables/tab_prior}

\subsection{Comparison with prior conversion methods}
\label{sec:res-prior}
Two earlier methods convert GazeBase toward the same VR device: percentile matching (PM; \citealp{griffith2025percentile}) and device-specific style transfer (ST; \citealp{lohr2025styletransfer}). We reimplemented PM and retrained ST, both on set~A only, and applied them to the GazeBase recordings. It is clear from Table~\ref{tab:prior} that, PM matches precision best, since precision is what it targets. Its noise, however, is white, unlike that of the VR device, so its sample-to-sample noise is about twice that of Set~B and a nearest-neighbour classifier almost always tells it from real data. Its accuracy also stays too good, because its per-fixation offsets have random signs and often cancel the recording's own error. ST regenerates the whole signal: its output is far too smooth and lands far from the targets. For authentication, PM falls below NDR in $d'$ at both lengths (NDR $-$ PM: $+0.51$ [$+0.26$, $+0.76$] at 45\,s and $+0.23$ [$+0.06$, $+0.39$] at 5\,s), and even below our parametric set, while the EERs of PM and NDR are close. ST does no better than plain resampling (Table~\ref{tab:auth}).

\subsection{Ablations}
\label{sec:res-ablation}
We evaluate the contribution of each component by removing or replacing one part of the NDR pipeline at a time in Table~\ref{tab:ablations}. Without Step~4 of the parametric device model, the Frequency-locked diffusion renderer starts from the signal after Step~3. This lowers $d'$ at both durations ($-0.90$ [$-1.10$, $-0.67$] at 45\,s and $-0.48$ [$-0.63$, $-0.33$] at 5\,s) and raises the EER by 3.1 [$+0.0$, $+5.9$] and 5.2 [$+2.0$, $+10.7$] percentage points. Trained only on noisy VR data and run with the small perturbation chosen in Section~\ref{sec:renderer}, the renderer can re-shape device noise but cannot create it; it needs the Step-4 noise as its starting point. Replacing the twin in Step~2 with the profile of a randomly drawn set A RAN recording lowers $d'$ at 5\,s ($-0.18$ [$-0.34$, $-0.04$]) and raises the EER by 3.2 points [$-0.0$, $+7.9$]. At 45\,s, its EER is lower than that of the NDR method (4.0 vs.\ 5.9\,\%), but the interval of this difference ($-1.9$ [$-4.8$, $+1.0$]) includes zero, and $d'$, which uses all scores, points the other way ($-0.14$ [$-0.36$, $+0.11$]). Finally, subtracting the recording's own EyeLink spectrum in Step~4, instead of the typical one, does not help. The EER barely changes ($0.0$ points at 45\,s and $-1.0$ at 5\,s; both intervals include zero), and $d'$ is slightly lower at 45\,s ($-0.15$ [$-0.32$, $+0.00$]).

\input{tables/tab_ablations}

%% file: tables/tab_quality.tex
\begin{table}[t]
  \caption{Signal quality and realism of the converted data. Medians over recordings, with the KS distance to Set B which indicated in parentheses. Set B is the 248 RAN recordings of the held-out GazeBaseVR.}
  \label{tab:quality}
  \footnotesize   
  \begin{tabular}{lcccc}
    \toprule
                          & Plain        & Parametric   & NDR                    & Set B \\
    \midrule
    Accuracy         & 0.860 (0.51) & 1.361 (0.06) & 1.358 (0.06)           & 1.333 \\
    MAD precision    & 0.086 (0.68) & 0.138 (0.25) & 0.129 (0.32)           & 0.166 \\
    RMS-S2S          & 0.029 (0.99) & 0.161 (0.16) & 0.142 (0.06)  & 0.147 \\
    STD              & 0.134 (0.60) & 0.193 (0.05) & 0.185 (0.11)           & 0.191 \\
    Signal type           & 0.208 (0.97) & 0.856 (0.24) & 0.788 (0.12)  & 0.770 \\
    Slope $\alpha$        & 2.273 (0.98) & 0.621 (0.12) & 0.685 (0.15)           & 0.641 \\
    Data loss (\%)        & 1.5 (0.49)   & 1.8 (0.53)   & 1.8 (0.53)             & 0.3 \\
    Realism, 1-NN (\%)    & 99.2         & 73.4         & 70.1                   & -- \\
    Person share, RMS-S2S & 0.68         & 0.78         & 0.80                   & 0.84 \\
    Person share, MAD     & 0.40         & 0.50         & 0.48                   & 0.72 \\
    \bottomrule
  \end{tabular}
\end{table}

%% file: tables/tab_auth.tex
\begin{table}[htbp]
  \caption{Authenticating GazeBaseVR set B users where enroll with session 1, verify with session 2.
  with EKYT trained on each conditions. Brackets: 95\,\% bootstrap CIs over test people.}
  \label{tab:auth}
  \footnotesize   
  \begin{tabular}{lcccccc}
    \toprule
                           & \multicolumn{2}{c}{5\,s}            & \multicolumn{4}{c}{45\,s} \\
    \cmidrule(lr){2-3} \cmidrule(lr){4-7}
    Training data          & EER (\%)         & $d'$             & EER (\%)        & $d'$             & FRR @ FAR $10^{-4}$ (\%) & Rank-1 (\%) \\
    \midrule
    Real VR        & 16.8 [13.4, 20.8] & 1.86 [1.66, 2.10] & 5.8 [3.2, 6.9]  & 3.10 [2.90, 3.32] & 56.4 & 86.1 \\
    \midrule
    Plain         & 33.7 [29.6, 37.6] & 0.81 [0.63, 1.03] & 12.8 [9.1, 16.8] & 1.76 [1.60, 1.95] & 96.0 & 67.3 \\
    Parametric & 23.7 [17.8, 27.7] & 1.46 [1.24, 1.71] & 6.9 [4.0, 9.9]  & 2.95 [2.73, 3.20] & 79.2 & 82.2 \\
    NDR             & 22.5 [17.8, 25.7] & 1.47 [1.24, 1.74] & 5.9 [3.0, 8.9]  & 3.21 [2.97, 3.48] & 64.4 & 80.2 \\
    \bottomrule
  \end{tabular}
\end{table}

%% file: tables/tab_prior.tex
\begin{table}[htbp]
\caption{Comparison of prior conversion methods and NDR against Set B. Signal quality: medians and KS distances. Authentication: EKYT performance on Set B; \(^{\dagger}\) trained on Set A data. PM: percentile matching \citep{griffith2025percentile}; ST: device-specific style transfer \citep{lohr2025styletransfer}; both fitted to Set A only.}
  \label{tab:prior}
  \footnotesize
  \begin{tabular}{lcccc}
    \toprule
                                  & \multicolumn{2}{c}{Prior work}        & Ours                          & Reference \\
    \cmidrule(lr){2-3} \cmidrule(lr){4-4} \cmidrule(lr){5-5}
                                  & PM                    & ST            & NDR                           & Set B \\
    \midrule
    \multicolumn{5}{l}{\textit{Signal quality}} \\
    MAD precision            & 0.178 (0.11) & 0.658 (0.99)  & 0.129 (0.32)                  & 0.166 \\
    RMS-S2S                  & 0.302 (0.75)          & 0.061 (0.97)  & 0.142 (0.06)         & 0.147 \\
    Signal type                   & 1.214 (0.93)          & 0.056 (1.00)  & 0.788 (0.12)         & 0.770 \\
    Accuracy                 & 1.042 (0.34)          & 7.979 (1.00)  & 1.358 (0.06)         & 1.333 \\
    Realism, 1-NN (\%)            & 98.6                  & 100.0         & 70.1                 & --    \\
    \midrule
    \multicolumn{5}{l}{\textit{Authentication of Set B}} \\
    $d'$, 45\,s / 5\,s            & 2.70 / 1.25           & 1.80 / 0.90   & 3.21 / 1.47 & 3.10 / 1.86$^\dagger$ \\
    EER (\%), 45\,s / 5\,s        & 6.0 / 22.8            & 11.9 / 31.7   & 5.9 / 22.5                    & 5.8 / 16.8$^\dagger$ \\
    \bottomrule
  \end{tabular}
\end{table}

%% file: tables/tab_ablations.tex
\begin{table}[htbp]
  \caption{Ablations of NDR, the parametric device model followed by the Frequency-locked diffusion renderer. Each row changes one step of the parametric device model and keeps the renderer; EKYT is trained on the result and tested on set B. Brackets: 95\,\% bootstrap CIs}
  \label{tab:ablations}
  \footnotesize
  \setlength{\tabcolsep}{5pt}
  \begin{tabular}{p{7.5cm}cccc}
    \toprule
                                         & \multicolumn{2}{c}{5\,s}                & \multicolumn{2}{c}{45\,s} \\
    \cmidrule(lr){2-3} \cmidrule(lr){4-5}
    Variant                              & EER (\%)          & $d'$              & EER (\%)        & $d'$ \\
    \midrule
    NDR: parametric device model + Frequency-locked diffusion renderer
                                         & 22.5 [17.8, 25.7] & 1.47 [1.24, 1.74] & 5.9 [3.0, 8.9]  & 3.21 [2.97, 3.48] \\
    \midrule
    \multicolumn{5}{l}{\textit{Change in the parametric device model}} \\
    \quad Step~2: random profile per recording, not the twin
                                         & 25.7 [21.0, 30.7] & 1.29 [1.10, 1.50] & 4.0 [2.1, 5.9]  & 3.07 [2.86, 3.33] \\
    \quad Step~4: own EyeLink spectrum, not the typical one
                                         & 21.6 [17.9, 24.8] & 1.51 [1.30, 1.76] & 5.9 [3.0, 8.9]  & 3.06 [2.79, 3.36] \\
    \quad No Step~4: Frequency-locked diffusion renderer starts after Step~3
                                         & 27.7 [24.2, 32.7] & 0.99 [0.79, 1.23] & 9.0 [5.9, 11.9] & 2.31 [2.11, 2.53] \\
    \bottomrule
  \end{tabular}
\end{table}

%% file: sections/07_discussion.tex
\section{Discussion}
\label{sec:discussion}

We asked whether lab recordings can be converted to the signal quality of a VR device while keeping their eye movement. Our conversion, NDR(the parametric device model followed by the frequency-locked diffusion renderer)  treats a recording as the eye movement plus what the device adds and re-creates only the device terms. The converted data match the VR device in noise level, signal type and accuracy (Table~\ref{tab:quality}), and a nearest-neighbour classifier separates them from real VR data far less often than plain lab data only. The eye movement itself also kept: saccade latency, the number of fixations and the main sequence remain those of the source (Figure~\ref{fig:content}). Priors methods such as Percentile matching \citep{griffith2025percentile} also keeps the eye movement, but its white noise leaves it easy to tell from real VR data, and style transfer \citep{lohr2025styletransfer} achieves neither (Table~\ref{tab:prior}). In contrast NDR also carries these signal-level gains over to the downstream test. With 45\,s of data, an authenticator trained only on converted recordings reaches a $d'$ close to that of real VR training (3.21 vs.\ 3.10) and a similar EER (5.9 vs.\ 5.8\,\%). The parametric device model brings most of this gain, and the renderer adds a smaller part. Although, With 5\,s of data, every converted set stays below real VR training ($d'$ 1.47 vs.\ 1.86 for NDR)

Several properties, however, remain off, and they show what the device model still misses. Converted gaze is too precise (MAD 0.13\textdegree{} vs.\ 0.17\textdegree): it lacks the slow drift of the VR signal. Converted saccades are faster than real VR saccades in every amplitude bin (Figure~\ref{fig:content}b) suggesting that the current model does not capture device-induced temporal filtering that may attenuate peak velocities. The model also uses stationary noise, independent of time, gaze position and eye speed, and preserves the missing-data gaps of the source recordings. These differences may help explain the performance gap at 5,s. The conversion adds VR-like noise but does not remove fine-scale eye-movement detail that the VR device may blur, potentially allowing an authenticator trained on converted data to exploit cues absent from real VR recordings. Differences in participant populations between the two datasets may also contribute to the gap. Closing these signal-level gaps while preserving the properties already matched remains a key challenge.

The ablation results also reveal trade-offs in the twin assignment and frequency-locking design. The twin keeps each converted person consistent across sessions, as real VR users are (person share of the noise level 0.80, against 0.00 with random profiles and 0.84 for set B), so its clear effect is on realism. Its effect on authentication is mixed (Section~\ref{sec:res-ablation}). It helps with 5\,s of data ($d'$ $+0.18$ over random profiles), but at 45\,s random profiles reach a lower EER (4.0 vs.\ 5.9\,\%), although this difference includes zero and $d'$ points the other way. This suggests that consistent device noise may itself act as an identity cue. The assumption that relative device quality transfers across devices remains untested; participants recorded on both devices, such as those shared by GazeBaseVR and SynchronEyes \citep{lohr2023gazebasevr,aziz2022synchroneyes}, could help test it. The frequency lock preserves source eye movements by restricting renderer changes to higher frequencies, but also prevents correction of the low-frequency precision gap.

Our study has several limitations. Each condition was trained once, so confidence intervals capture test-participant uncertainty but not training randomness. Training-data sizes also differ, and the renderer uses all five Set~A tasks, qualifying comparisons with real-VR training. Authentication covers only same-day sessions, not long-term performance. We studied one device pair and task; the twin and renderer require unpaired VR recordings, and their minimum data requirements remain unknown. Person-specific device noise may itself act as an identity cue: converted participants inherit their twin's noise signature, with 34 GazeBase participants sharing a twin with another participant. EKYT may therefore exploit noise rather than eye movement, an effect we did not quantify. Finally, we did not test sensitivity to the chosen frequency $f^*$. Future work will be longer enrolment--probe intervals and people recorded on both devices, and test whether converted recordings keep the identity of their source. The device model can be made richer, for example with noise that changes over time and a device filter. Finally, other headsets, tasks and downstream models, and the combination of converted with real VR training data, would show how far the approach generalises and what it is worth in practice.

%% file: sections/08_conclusion.tex
\section{Conclusion}
\label{sec:conclusion}
We converted lab eye-tracking recordings to the signal quality of a VR headset while keeping their eye movement. Treating a VR recording as the eye movement plus what the device adds, NDR re-creates only the device: camera sampling, the calibration error and spectrally shaped noise of a matched VR twin, and a diffusion renderer that re-textures the noise and leaves the signal below $f^*$ essentially unchanged. Converted GazeBase recordings match GazeBaseVR in noise level, signal type and accuracy, are much harder to tell from real VR data than resampled lab data, and keep the saccade latency and dynamics of their source. In a downstream test, an authenticator trained only on converted recordings reached, with 45\,s of data, a $d'$ on unseen real VR users close to that of one trained on real VR data (3.21 vs.\ 3.10), while with 5\,s of data it stays below real VR training (EER 22.5 vs.\ 16.8\,\%). Converted gaze is still too precise, its saccades are faster than those of the VR device, and its noise does not change over time. Code and split lists will be released upon publication.

%% file: sections/09_statements.tex

\section*{Privacy and Ethics Statement}
This work uses only two public datasets, GazeBase and GazeBaseVR, whose participants gave informed consent to the collection and de-identified sharing of their data, and which are distributed under the CC BY 4.0 license \citep{griffith2021gazebase,lohr2023gazebasevr}; no new data were collected.

\section*{Use of Large Language Models}
The authors used a large language model (Claude, Anthropic) as an assistant for refactoring and documenting code, drafting and editing the manuscript text, and checking the bibliography against Crossref and arXiv records. The authors verified all analyses, numbers and statements and take full responsibility for the content of this paper.